# A Versatile Apparatus for Measuring the Growth Rates of Small Ice Prisms from the Vapor Phase

Kenneth G. Libbrecht

Department of Physics
California Institute of Technology
Pasadena, California 91125
kgl@caltech.edu

**Abstract.** I describe an adaptable apparatus for making precision measurements of the growth of faceted ice prisms from water vapor as a function of temperature, supersaturation, and background gas pressure. I also describe procedures for modeling growth data to disentangle a variety of physical effects and better understand systematic errors and measurement uncertainties. By enabling precise ice-growth measurements over a broad range of environmental conditions, this apparatus is well suited for investigating the molecular attachment kinetics at the ice/vapor interface, which is needed to understand and model snow crystal growth dynamics.

# 1. Snow Crystal Attachment Kinetics

While the formation of atmospheric snow crystals is a common natural phenomenon, understanding the physical origins of observed snow crystal morphologies has proven to be a remarkably difficult task [2019Lib, 2017Lib]. In typical environmental conditions, the development of snow crystal structure is governed by two dominant processes: particle diffusion of water vapor through the surrounding air and the molecular attachment kinetics at the ice surface. Lesser considerations include the diffusion of latent heat, surface-energy factors (notably the Gibbs-Thomson effect), and perhaps other factors that can become quite important in unusual circumstances [2019Lib].

Much of this underlying physics is well understood and calculable at a fundamental level. For example, particle and heat diffusion involve straightforward physical principles and can be readily incorporated into finite-element computational models. The ice surface energy is not so well characterized, especially the surface-energy anisotropy, but this is a small effect and it appears that a simple isotropic surface-energy model is sufficient for understanding snow crystal growth [2012Lib2, 2019Lib].

The ice/vapor attachment kinetics, on the other hand, is quite an intricate and puzzling phenomenon [2019Lib1, 1987Kob], involving subtle molecular dynamics processes taking place at a structurally complex ice/vapor interface. Thus, while much progress has been made toward developing computational models of faceted diffusion-limited growth [2009Gra, 2014Kel], the creation of a fully functional snow-crystal simulator awaits a better comprehension of the physical underpinnings of the ice/vapor attachment kinetics.

In this paper I describe a relatively simple experimental apparatus that allows precise measurements of ice growth rates as a function of temperature, supersaturation, background gas pressure, and surface orientation. Previous measurements of this nature have allowed significant progress toward understanding the attachment kinetics [2013Lib], and further progress on this experimental front is required to test new physical models of the relevant molecular processes [2019Lib1].

It is customary to describe the ice/vapor attachment kinetics in terms of the Hertz-Knudsen relation [1882Her, 1915Knu, 1996Sai, 1990Yok, 2005Lib, 2017Lib, 2019Lib]

$$v_n = \alpha v_{kin} \sigma_{surf} \quad (1)$$

where $v_n$ is the crystal growth velocity normal to a growing surface, $\alpha$ is a dimensionless *attachment coefficient*, $\sigma_{surf}$ is the water vapor supersaturation at the surface, and the *kinetic velocity* $v_{kin}$ incorporates the statistical mechanics of ideal gases. A detailed discussion of this equation and its foundations can be found in [2019Lib].

For most lattice orientations of the ice surface, $\alpha \approx 1$ is a reasonable approximation, but this is not the case on the principal basal and prism facets. Indeed, the anisotropy of the attachment kinetics ($\alpha_{basal}, \alpha_{prism} \ll 1$) is ultimately responsible for the formation of these faceted surfaces [2019Lib]. Much of the behavior of $v_{basal}$ and $v_{prism}$ can be explained by the nucleation and growth of 2D terraces on the facet surfaces, which is related to the terrace step energies via classical nucleation theory [2019Lib, 2013Lib, 1996Sai, 1998Nel].

Terrace nucleation is far from the whole story, however, as there is now substantial evidence suggesting that the nucleation rate depends strongly on the size of the facet surfaces, as the nucleation process is substantially affected by nearby corner structures [2019Lib1, 2019Lib2]. The resulting

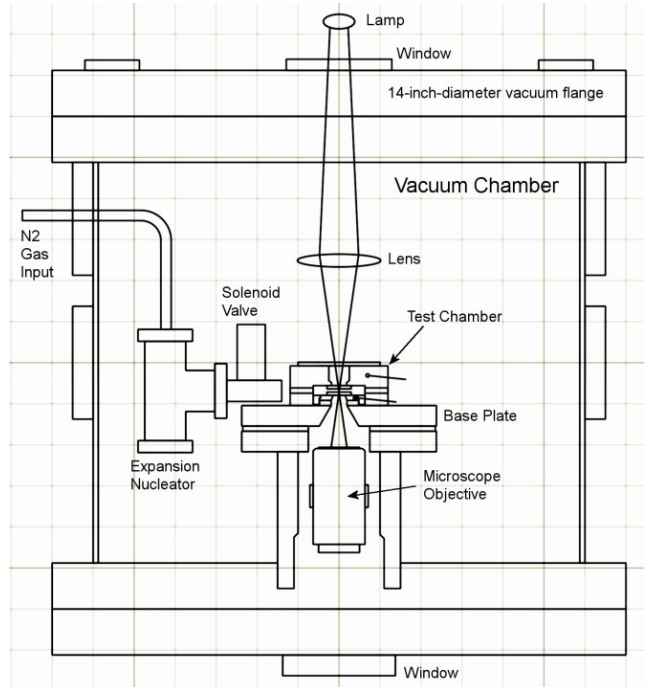

**Figure 1: A sketch of the main components of the apparatus described in this paper. The cooled vacuum chamber defines the experimental space, and the scale of the sketch is set by the 14-inch vacuum flanges. The expansion nucleator [2019Lib] creates an on-demand source of small test crystals, and the microscope objective focuses the substrate onto a 40-megapixel camera outside the vacuum envelope.**

*Structure Dependent Attachment Kinetics* (SDAK) [2003Lib1, 2019Lib1] appears to be responsible for many of the peculiar growth behaviors that have thwarted efforts to understand snow crystal growth dynamics for many decades. With a newly proposed physical model of the SDAK process [2019Lib1], a current goal is to test this hypothesis over a broad range of conditions using targeted growth experiments, and some results provided by the apparatus described here can be found in [2019Lib2].

## 2. A Variable-Pressure Growth Chamber

Figure 1 shows a sketch of this new apparatus, which was designed to make precise measurements of the growth of small ice crystals under carefully controlled environmental conditions. The outer vacuum envelope is cooled using a recirculating chiller [2019Lib] that provides primary cooling for the entire apparatus, capable of reaching temperatures down to -35 C. Copper plates with soldered copper tubing attach to the top and bottom vacuum flanges as well as around the body of the vacuum chamber. These cooling plates, covered with styrofoam insulation, are not shown in Figure 1.

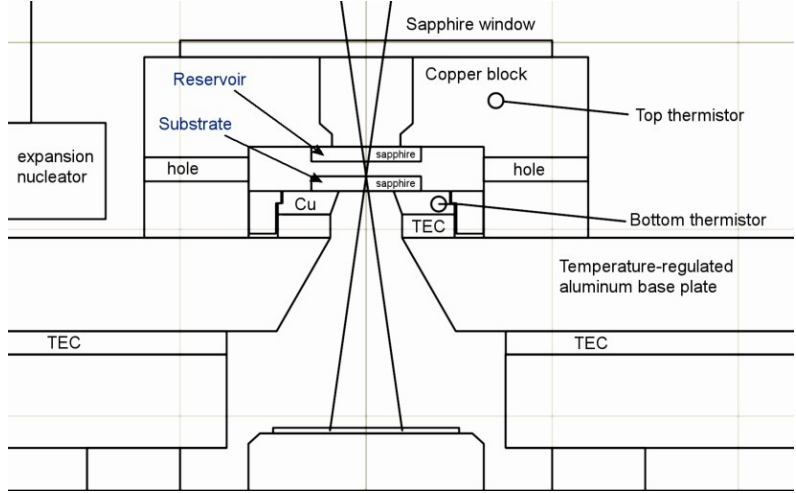

Figure 2: A detailed look at the test chamber shown in Figure 1. The bottom surface of the reservoir plate is typically covered with a thin layer of ice, providing a source of water vapor for small ice crystals growing on the upper surface of the substrate. This yields the parallel-plate geometry of the inner test chamber illustrated in Figure 3.

The temperature of the base plate shown in Figure 1 is controlled to an absolute accuracy of better than 0.1 C using a precision thermistor and thermal-electric modules within the vacuum envelope. The small test chamber is coupled to the base plate using thermal joint compound, and this copper block provides a thermal anchor point for the test assembly shown in Figure 2.

Lifting the copper block from the base plate in Figure 2 allows access to the ice reservoir and substrate surfaces, which are thoroughly cleaned between runs. After cleaning, usually a hydrophobic surface coating is applied to the substrate surface to reduce possible substrate interactions. I have been obtaining satisfactory results using *Hendlex Nano Glass Pro and Glass Prepare Cleaner* on the sapphire substrate, but it is difficult to ascertain the coating effectiveness at low temperatures. Whenever ice crystals are grown on a substrate, as they are in this apparatus, substrate interactions must be considered as a possible source of systematic errors in the growth measurements.

One method to load ice onto the reservoir surface is to lay a wet Kimwipe tissue over this and adjacent surfaces right before cooldown. The tissue adheres to the reservoir surface and turns to ice during cooldown, providing a large supply of ice. The tissue is sufficiently translucent to allow illumination for imaging the substrate, as shown in Figure 1. Another method is to leave the reservoir surface dry during cooldown and then blow warm room air onto the surface while heating the substrate. Ice crystals readily form on the reservoir this way, and imaging the reservoir surface instead of the substrate allows monitoring of the ice content on the reservoir.

During operation, the substrate is temperature regulated using a specially designed electronic controller that fixes the temperature difference between the copper block and the substrate. This circuit also outputs a monitor voltage equal to

$$V_{mon} = G \frac{R_{T,top}}{R_{T,top} + R_{T,bottom}} V_{in} \quad (2)$$

where $R_{T,top}$ and $R_{T,bottom}$ are the resistances of the top and bottom thermistors in Figure 2, with $G$ and $V_{in}$ provided by precision electronic components. As described in the analysis section below, $V_{mon}$ is related to the temperature difference between the two plates, and thus to the supersaturation at the substrate surface.

The expansion nucleator shown in Figure 1 consists of a small vacuum tee connected to a source of humidified compressed nitrogen gas and a solenoid valve [2019Lib]. When the valve is opened, the compressed gas rapidly expands and cools, nucleating minute ice crystals in the process. Some of these crystals spray into the test chamber and land on the waiting substrate surface, where their subsequent growth can then be monitored by the optical imaging system.

Note that the diffusion timescale between the sapphire plates in Figure 2 is roughly $\tau \approx L^2/D$, where $L$ is the spacing between the plates and $D$ is the diffusion constant. After a nucleation pulse in air at a pressure of one bar, therefore, the supersaturation field between the plates settles to its quasi-equilibrium state after a time of about $\tau \approx 0.2$ seconds. At lower pressures, this time is substantially reduced.

The overall aim of this apparatus is to produce the simple plane-parallel geometry illustrated in Figure 3. The reservoir surface at temperature $T_{reservoir}$ provides an essentially infinite source of water vapor, which is absorbed by the growing test crystals on the substrate at temperature $T_{substrate}$. No growth occurs when $T_{reservoir} = T_{substrate}$, and the supersaturation $\sigma_{subst}$ near the substrate is roughly proportional to the temperature difference $\Delta T = T_{reservoir} - T_{substrate}$, as is further described in the analysis section below.

## 3. Physical Processes in Ice Crystal Growth

To extract information about the attachment kinetics from observations of growing crystals, it is necessary to disentangle influences from particle diffusion, the Gibbs-Thomson effect, and other factors. While 3D computational modeling is required to fully understand the growth of complex snow-crystal morphologies, basic analytical techniques are remarkably useful for examining the growth of simple prisms. To develop these techniques,

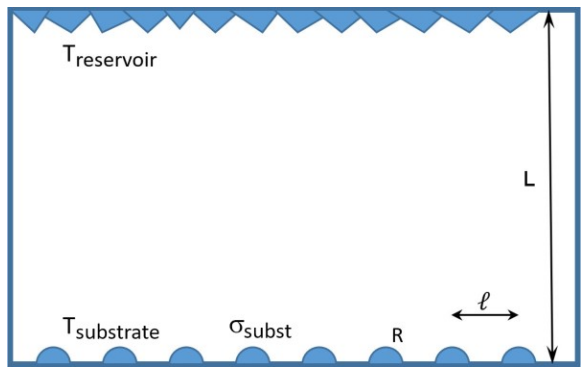

**Figure 3: An idealized schematic (not to scale) of a small section of the inner parallel-plate ice growth chamber. The top and bottom surfaces are provided by a pair of 15-mm-diameter, 2-mm-thick sapphire windows with a separation of L = 2 mm. The top surface is covered with a thick layer of frost crystals, providing an essentially infinite water-vapor reservoir. Small test crystals are deposited randomly on the lower surface, where their growth can be measuring using optical microcopy.**

first consider the growth of hemispherical ice crystals as illustrated in Figure 3.

### Spherical Analysis

Because the spacing between the reservoir and substrate surfaces is much less than their lateral extent, I assume periodic boundary conditions in the analysis, which is equivalent to assuming a plane-parallel geometry with infinite lateral extent. As shown in Figure 3, the top surface is kept at a fixed temperature $T_{reservoir}$, and this surface is covered with a thick layer of ice crystals that serves as a water-vapor reservoir.

Because the diffusion time scale $\tau$ is very short (the Laplace approximation [2019Lib]), and the ice reservoir is large, the water-vapor number density at the surface of the reservoir is always close to the equilibrium value $c_{reservoir} = c_{sat}(T_{reservoir})$, where $c_{sat}(T)$ is the normal saturated water vapor density above a flat ice surface at an equilibrium temperature $T$. This value thus provides a fixed upper boundary condition in our solution of the particle diffusion equation to determine the supersaturation surrounding the test crystals.

The lower surface has a fixed temperature $T_{substrate}$, and the heat diffusion equation yields the simple solution of a linear vertical temperature profile that goes from $T_{reservoir}$ at the top surface to $T_{substrate}$ at the bottom surface. In the absence of any test crystals, the solution to the particle diffusion equation then gives a constant number density $c = c_{sat}(T_{reservoir})$ throughout the entire region between the two plates. Thus, in the absence of any test crystals, the water vapor supersaturation at the lower plate is

$$\sigma_{subst,0} \approx \frac{1}{c_{sat}} \frac{dc_{sat}}{dT} \Delta T$$
$$\approx \eta \Delta T \quad (3)$$

where $\Delta T = T_{reservoir} - T_{substrate}$.

It is not trivial to determine the absolute value of $\Delta T$ with great precision in this apparatus, owing to small offsets in thermistor values at a fixed temperature. In anticipation of further analysis considerations described below, therefore, I rewrite Equation 3 in the experiment-friendly form

$$\sigma_{subst,0} = C_{cal}(V_{mon} - V_{mon,0}) \quad (4)$$

where

$$C_{cal} = \frac{4\eta}{GV_{in}} \left( \frac{1}{R_T} \frac{dR_T}{dT} \right)^{-1} \quad (5)$$

The value of $C_{cal}$ is known to a reasonably high precision, but $V_{mon,0}$ is a somewhat critical parameter that is best determined by examining the crystal growth behavior, as I describe in the modeling section below.

## Large-Scale Diffusion

The first extension of Equation 4 arises from the collective growth of all the crystals on the substrate, and I refer to this as a large-scale diffusion (LSD) correction. Referring to Figure 3, I model this system by assuming a uniform square array of growing hemispherical crystals, each having a radius $R$, each growing at the same velocity $v$, with a nearest-neighbor separation of $\ell$, so the areal crystal number density is $\ell^{-2}$. I assume that $R \ll \ell \ll L$, where $L = 2\ mm$ is the plate separation shown in Figure 3, and both these inequalities are valid for typical data sets, as can be verified by direct imaging.

The LSD correction arises from the overall downward flux of water vapor, in this model given by

$$F = \frac{dV}{dt} \frac{c_{ice}}{\ell^2} \quad (6)$$

where

$$\frac{dV}{dt} = 2\pi R^2 v \quad (7)$$

for a hemispherical crystal. From the diffusion equation, this flux must be driven by a gradient in the water-vapor density, thus yielding the corrected supersaturation

$$\sigma_{subst} \approx \sigma_{subst,0} - \delta\sigma_{LSD} \quad (8)$$

with

$$\delta\sigma_{LSD} \approx \frac{FL}{c_{sat}D}$$
$$\approx 2\pi \frac{Lv}{D} \frac{R^2}{\ell^2} \frac{c_{ice}}{c_{sat}} \quad (9)$$

where $D$ is the water-vapor diffusion constant, equal to $D \approx 2 \times 10^{-5}\ m^2/s$ in normal air at a pressure of one bar [2019Lib].

In practice, the LSD correction quantifies the question of how crystal crowding on the substrate affects the growth measurements. For a given data set, $\ell$ can be estimated simply by counting crystals, and the growth parameters are measured from a single test crystal. Usually the assumption of identical crystals is not unreasonable, and this too can be checked by examining the crystal-to-crystal uniformity. If Equation 9 yields a sufficiently low value $\delta\sigma_{LSD}$, then the correction is manageable, meaning that $\sigma_{subst}$ is fairly well determined surrounding the test crystals. If the value of $\delta\sigma_{LSD}$ is considered too high, however, then the data set must be rejected, suggesting that additional data with larger $\ell$ (less crystal crowding) must be obtained.

Note that $\delta\sigma_{LSD}$ is proportional to the spacing $L$, so minimizing the (often substantial) LSD correction suggests making $L$ as small as possible. This requirement is what drove the development of the parallel-plate geometry used in this apparatus [2013Lib].

## SMALL-SCALE DIFFUSION

Assuming that the LSD correction is manageable, then the value of $\sigma_{subst}$ at the substrate is fairly well known. Because $R \ll \ell \ll L$, we can then assume that $\sigma_{subst}$ is essentially equal to the supersaturation at distances far from a growing crystal, which one normally calls $\sigma_\infty$ in diffusion analyses. Continuing with our spherical analysis, the supersaturation $\sigma_{surf}$ at the growing crystal surface is then given by [2019Lib]

$$\sigma_{surf} = \sigma_{subst} - \delta\sigma_{SSD} \qquad (10)$$

where the small-scale diffusion correction is equal to

$$\delta\sigma_{SSD} = \frac{R}{X_0}\frac{v}{v_{kin}} \qquad (11)$$

and $X_0$ is the characteristic diffusion length [2019Lib]

$$X_0 = \frac{c_{sat}}{c_{ice}}\frac{D}{v_{kin}} \qquad (12)$$

For example, at -5 C in normal air, $X_0 \approx 0.142\ \mu m$ and $v_{kin} \approx 496\ \mu m/sec$, giving

$$\delta\sigma_{SSD} \approx 1.4\% \cdot \left(\frac{R}{10\mu}\right)\left(\frac{v}{0.1\ \mu m/sec}\right) \qquad (13)$$

This correction can be quite large under growth conditions in normal air, but it is also roughly proportional to the background gas pressure. Thus both the LSD and SSD corrections can be greatly reduced by operating at low pressures.

Note that this result follows from solving the particle diffusion equation around a growing spherical crystal, which also yields

$$v = \frac{\alpha\alpha_{diff}}{\alpha + \alpha_{diff}}v_{kin}\sigma_\infty \qquad (14)$$

where $\alpha_{diff} = X_0/R$ [2019Lib].

The spherical analysis is especially useful for determining whether one is in a regime where the diffusion corrections are small enough to obtain useful information about the attachment kinetics. An optimal situation is one where $\alpha \ll \alpha_{diff}$, which is equivalent to $\delta\sigma_{SSD} \ll \sigma_{subst}$.

Note that the separation of diffusion corrections into large-scale and small-scale terms is a reasonable division because of the assumed inequalities $R \ll \ell \ll L$, which can be easily verified for each data set. In this picture, the collective diffusion contributions from all the crystals overlap and combine at length scales of order $\ell$ and above, yielding the near-substrate supersaturation $\sigma_{subst} < \sigma_{subst,0}$. In addition, each crystal produces its own isolated "hole" in the supersaturation field with a size of order $R$, which gives $\sigma_{surf} < \sigma_{subst}$.

## THE GIBBS-THOMSON EFFECT

The apparatus described here was designed to observe exceptionally small ice crystals growing at exceedingly low supersaturations, and under such conditions surface-energy effects can be important. Most significant is the Gibbs-Thomson effect, which replaces Equation 1 with [2019Lib]

$$v_n = \alpha(\sigma_{surf})v_{kin}(\sigma_{surf} - d_{sv}\kappa) \qquad (15)$$

where $d_{sv} \approx 1.0\ nm$ and $\kappa$ is the local surface curvature, equal to $\kappa = 2/R$ for a spherical surface. This expression can be rewritten as

$$v_n = \alpha v_{kin}\frac{\alpha_{diff}}{\alpha + \alpha_{diff}}(\sigma_\infty - d_{sv}\kappa) \qquad (16)$$

where here $\alpha = \alpha(\sigma_{surf})$ and $\sigma_\infty = \sigma_{subst}$ is appropriate for this apparatus, as described above. Rearranging these expressions then gives

$$\sigma_{surf} = \frac{\alpha_{diff}\sigma_\infty + \alpha d_{sv}\kappa}{\alpha + \alpha_{diff}} \qquad (17)$$

The value of $d_{sv}\kappa$ is usually relatively small, providing a significant correction for only the smallest crystals growing at especially low supersaturations. The Gibbs-Thomson effect is most noticeable immediately after nucleation, as the expansion nucleator produces crystals that are typically of order one micron in size. If $\sigma_{subst,0}$ is too low, these nascent crystals quickly sublimate away. Thus the Gibbs-Thomson effect tends to set a practical lower limit to $\sigma_{subst,0}$, below which it is not possible to nucleate crystals.

## Substrate Interactions

Another potential systematic error in ice growth measurements arises from substrate interactions. Of particular importance is the heterogeneous nucleation of new molecular terraces on faceted ice surfaces that intersect the substrate, as is described in [2019Lib, 2012Lib]. In [2013Lib], we avoided this problem by restricting our observations to facets that did not contact the substrate, but that method precluded using simultaneous measurements of many crystals to obtain a greater experimental throughput.

After additional investigation, it now appears that substrate interactions are not a huge problem, so hydrophobic coatings are applied to the substrate surface in the present apparatus to increase the ice/substrate contact angle and hopefully minimize substrate influences on the measured crystal growth rates. It appears that this strategy works fairly well except at exceptionally low growth rates [2019Lib2], but additional investigation is needed to examine this issue further.

## Chemical Impurities

It is a practical impossibility to completely eliminate chemical impurities from any experimental apparatus, and it is difficult to know what impurity level is required before chemical effects are negligible. The available evidence suggests, however, that fairly high levels are needed to effect large changes in growth rates relative to perfectly clean air or a near-vacuum environment [2019Lib]. In the present apparatus, the test chamber is opened and cleaned between each run, and it appears that this is sufficient to reduce residual chemical effects to an acceptable level.

## Latent Heating

Another factor that can affect ice growth is the thermal diffusion that removes latent heat generated by solidification. When a test crystal is resting on a substrate, this heat is mostly conducted through the ice to the substrate below, which acts as an infinite heat reservoir. This heat flow can be approximated by considering an infinite sheet of ice of thickness $R$ resting on the substrate. If the sheet grows upward with a velocity $v$, then latent heating produces a temperature increase

$$\delta T \approx \frac{L_{sv}\rho_{ice}vR}{\kappa_{ice}} \qquad (18)$$

where $L_{sv}$ is the solid/vapor latent heat per unit mass, $\rho_{ice}$ is the density of ice, and $\kappa_{ice}$ is the thermal conductivity of ice. This yields an effective supersaturation correction of

$$\delta\sigma_{therm} \approx \frac{\eta L_{sv}\rho_{ice}vR}{\kappa_{ice}} \qquad (19)$$

For example, at -5 C, $\eta \approx 0.082\,K^{-1}$, $L_{sv} \approx 2.8 \times 10^6\,J/kg$, $\rho_{ice} \approx 917\,kg/m^3$, and $\kappa_{ice} \approx 2.3\,Wm^{-1}K^{-1}$, giving

$$\delta\sigma_{therm} \approx 0.01\% \cdot \left(\frac{R}{10\mu}\right)\left(\frac{v}{0.1\,\mu m/s}\right) \qquad (20)$$

For hemispherical ice prisms as illustrated in Figure 3, one must add a dimensionless geometrical factor to this result that cannot be calculated analytically, but I estimate that this increases $\delta\sigma_{therm}$ by roughly a factor of two.

Comparing $\delta\sigma_{therm}$ in Equation 20 to $\delta\sigma_{SSD}$ in Equation 13, we see that the thermal

correction is often negligible compared to particle diffusion effects. However, $\delta\sigma_{therm}$ cannot be reduced by operating at a low background gas pressure, so this can become an important correction factor at low pressures.

Another consideration when calculating $\delta\sigma_{therm}$ concerns the nature of the ice/substrate thermal coupling. If an ice crystal facet rests flat against a bare sapphire substrate, then the thermal surface coupling is likely quite good, so the above analysis of $\delta\sigma_{therm}$ will be reasonably accurate. The addition of even a thin hydrophobic coating may change this analysis significantly, however, if the thermal conductivity of the coating is substantially less than that of ice. Moreover, superhydrophobic coatings may result in much larger thermal corrections, as they are often quite thick, involving minimal surface contact and internal microscopic air gaps.

The addition of surface coatings may therefore introduce large and difficult-to-calculate thermal corrections. Thus, while hydrophobic and superhydrophobic coatings can decrease unwanted substrate interactions, they may also increase unwanted thermal effects, so one should proceed with caution when using such coatings.

## Simple Hexagonal Prisms

When using the apparatus described above, it is typical to focus on well-formed simple hexagonal prisms with one prism facet resting flat against the substrate. This allows measurement of the effective radius $R$ of the prism (approximating the hexagonal cross section by a simple circle), the half-height $H$ of the prism, and the growth velocities $v_R = dR/dt$ and $v_H = dH/dt$. For such crystals, it is then possible to obtain information about both $\alpha_{basal}$ and $\alpha_{prism}$ from a single series of measurements.

In contrast to the spherical case, it is not possible to produce a simple analytical analysis of the growth of faceted ice prisms [2001Woo]. And while a full 3D numerical analysis is possible, this can become quite tedious when examining the growth of many crystals. A 2D cylindrically symmetrical numerical analysis is simpler and faster, but again somewhat laborious in practice.

To expedite the analysis of crystal growth data taken with this apparatus, therefore, I have adapted the spherical analysis described above for application to the growth of simple hexagonal prisms. The resulting "1.5D" model includes separate $\alpha_{basal}$ and $\alpha_{prism}$ terms, but uses the monopole approximation described in [2019Lib] to treat diffusion effects. As I describe in the following section, this simple-prism model allows rapid numerical results, inclusion of the full range of correction factors described above, and provides a straightforward way to examine measurement and modeling uncertainties with relative ease.

To apply the monopole approximation, I define a volume-conserving effective spherical radius

$$R_{eff} = \left(\frac{2}{3}R^2 H\right)^{1/3} \quad (21)$$

so the volume of a sphere of this radius equals the total volume of the corresponding hexagonal prism.

In addition, because the rate of change of the volume of a hexagonal prism is equal to

$$\frac{dV}{dt} = 4\pi R\left(Hv_R + \frac{Rv_H}{2}\right)$$
$$\approx 4\pi R v_{kin}\sigma_{surf}\left(H\alpha_{prism} + \frac{R\alpha_{basal}}{2}\right) \quad (22)$$

I define an effective attachment coefficient

$$\alpha_{eff} = \left(H\alpha_{prism} + \frac{R\alpha_{basal}}{2}\right)\frac{R}{R_{eff}^2} \quad (23)$$

which defines an effective spherical crystal having the same total growth rate (the same $dV/dt$) as the hexagonal prism. How these quantities are put into use is described in the next section.

I also modify the Gibbs-Thomson effect to deal with the hexagonal-prism case to a reasonable approximation by using the effective surface curvatures

$$\kappa_{basal} \approx \frac{2}{R}$$
$$\kappa_{prism} \approx \frac{1}{R} + \frac{1}{H} \quad (24)$$

For the case of plate-like ice prisms, for example, this formulation provides that the thin prism edges will begin sublimating sooner than the broad basal faces when $\sigma_{surf}$ is slowly reduced to negative values, as is observed in the data example presented below.

In typical use, the 1.5D model provides a convenient and reasonably accurate method for modeling the growth of simple ice prisms, as long as the correction factors are not too large. In the $\alpha \ll \alpha_{diff}$ limit, the diffusion corrections are small and the attachment coefficients $\alpha_{basal}$ and $\alpha_{prism}$ can easily be extracted from the growth data using Equation 1. The purpose of the 1.5D model is to allow measurement of both $\alpha_{basal}$ and $\alpha_{prism}$ even when $\alpha \approx \alpha_{diff}$. The model works best for nearly isometric prisms with $R \approx H$, and will give somewhat distorted results for thin plates or slender columns. Of course, when $\alpha \gg \alpha_{diff}$, it becomes exceedingly difficult to glean much useful information about the attachment coefficients from growth measurements [2019Lib, 2019Lib2].

## 4. Forward Modeling of Hexagonal-Prism Growth

A typical series of ice-growth data consists of a set of time-stamped images of the substrate with a corresponding measurement of the monitor voltage $V_{mon}(t)$. From a visual inspection of the images, individual crystals are selected for detailed analysis, typically by finding especially well-formed prisms that are oriented with one prism facet resting flat against the sapphire surface. Measurements made from the images then yield $R(t)$ and $H(t)$ for each selected crystal. Often several suitable crystals are analyzed to examine crystal-to-crystal variation within a single run, and multiple runs using ostensibly identical growth conditions are examined as well.

Because the diffusion corrections and other factors described in the preceding sections are often quite significant, I have found it useful to compare the measured $R(t)$ and $H(t)$ with model calculation of $R_{mod}(t)$ and $H_{mod}(t)$ that are generated in the 1.5D prism model alluded to previously. In this section I describe the model calculations in some detail.

I follow a standard forward-modeling strategy in which a small seed crystal is numerically grown using a set of input initial conditions and physical parameters. The model inputs are then adjusted to reproduce the data, hopefully providing a reasonable fit to the entire set of $R(t)$ and $H(t)$ measurements. Importantly, this modeling strategy allows one to adjust each of the parameters to different degrees, and this process usually facilitates a satisfactory understanding of the measurement uncertainties and possible systematic errors. Model inputs include:

• The measured $V_{mon}(t)$ for the series, which usually exhibits a pre-determined time-dependent behavior. Often $V_{mon}(t)$ is kept constant for a time after nucleation to allow for growth under constant conditions, and then $V_{mon}(t)$ is ramped to slowly lower the supersaturation until sublimation occurs.

• The value of $V_{mon,0}$ that defines the point at which the supersaturation $\sigma_{subst,0}$ goes to zero. This value can be measured from independent measurements, but not to high accuracy. Therefore, $V_{mon,0}$ is usually treated as an adjustable parameter determined mainly from the onset of sublimation in the image data, for the specific test crystal being analyzed.

• The attachment coefficients $\alpha_{basal}(\sigma_{surf})$ and $\alpha_{prism}(\sigma_{surf})$, usually derived from a pre-determined functional form with a small number of input parameters.

- Initial values of the crystal size, $R(t = 0)$ and $H(t = 0)$.

Starting from these inputs, the code then calculates the growth of the crystal following these computational steps:

- Calculate $\sigma_{subst,0}$ using Equation 4.

- Compute $\sigma_{subst} = \sigma_{subst,0} - \delta\sigma_{LSD}$ using Equation 8 and the value of $\delta\sigma_{LSD}$ determined below. For the first iteration of the growth series, both $\sigma_{subst}$ and $\sigma_{surf}$ are set equal to $\sigma_{subst,0}$.

- Calculate $\sigma_{surf}$ using an iterative algorithm. From an initial estimate of $\sigma_{surf}$, calculate $\alpha_{eff}(\sigma_{surf})$ using Equation 23. Then calculate $\sigma_{surf,next}$ using Equation 17 with $R_{eff}$ provided by Equation 21, set $\sigma_{surf} \to 0.9\sigma_{surf} + 0.1\sigma_{surf,next}$, and iterate. As long as the diffusion corrections are not too large, this process usually converges quickly, yielding $\sigma_{surf}$, $\alpha_{eff}(\sigma_{surf})$, $\alpha_{basal}(\sigma_{surf})$ and $\alpha_{prism}(\sigma_{surf})$. This iterative algorithm uses the monopole (spherically symmetric) approximation to estimate the supersaturation at $R_{eff}$, which is then used as an estimate of $\sigma_{surf}$ at all points on the crystal surface. Because this formulation conserves the crystal volume $V$ and the volume growth $dV/dt$, as described above, it typically yields a reasonable approximation of $\sigma_{surf}$ for roughly isometric prisms.

- Calculate the growth velocities for both the basal and prism facets using Equations 16 and 24, and apply $H \to H + v_{basal}dt$ and $R \to R + v_{prism}dt$ to grow the crystal.

## A Modeling Example

To illustrate the forward modeling process on some real data, Figure 4 shows a section of an image of several ice crystals growing on the substrate. These crystals were grown at a temperature of -5 C in a reduced air pressure of 0.08 bar. After growing at an initial supersaturation of $\sigma_{subst,0} = 0.28\%$ for about 100 seconds, the supersaturation was slowly lowered in a linear ramp until the crystals showed significant sublimation. The circled crystal in Figure 4 was selected for further analysis, and Figure 5 shows a series of images of this crystal as it grew and then sublimated. Figure 6 shows measurements of $R(t)$ and $H(t)$ extracted from these 27 images.

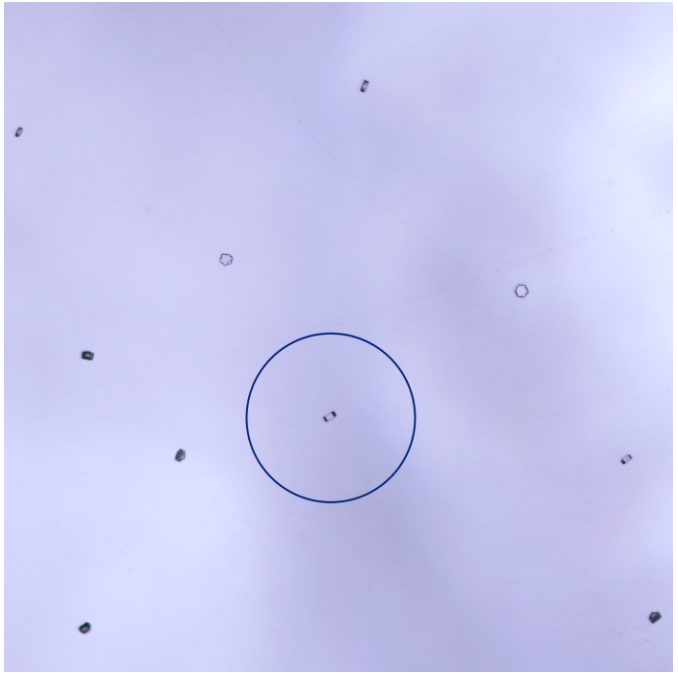

**Figure 4: A 1mm x 1mm portion of a single 4mm x 5mm image, taken 150 seconds after a nucleation pulse deposited about 150 tiny ice crystals on the substrate. The background illumination has a mottled appearance because the illumination source passes through the ice reservoir located 2mm above the substrate. From a visual inspection of the entire field, the circled crystal was selected for further analysis shown in Figures 5 and 6.**

Referring to the black supersaturation line in Figure 6, this shows $\sigma_{subst,0}$ derived from the measured $V_{mon}(t) - V_{mon,0}$ that was recorded as the crystals were growing, using Equation 4. The lower $\sigma_{subst}$ and $\sigma_{surf}$ curves were derived from the model, showing the magnitude of the two diffusion corrections. These corrections are generally small at low growth rates and low pressures, but can be quite large at an air pressure of 1 bar.

The initial spike in $\sigma_{subst,0}$ seen in Figure 6 arises from the pulse of cold air exiting the expansion nucleator. This air cools the substrate plate slightly, thus increasing $\sigma_{subst,0}$ by a small amount, and it takes some time for the temperature controller to correct this perturbation. The reservoir plate is also cooled by this pulse of cold air, but it is strongly coupled to the copper block that supports it, and thus the reservoir plate re-equilibrates in a short time. The substrate plate is only weakly thermally coupled via the temperature servo, however, which responds relatively slowly. The supersaturation ramp after $t = 90$ seconds was driven by changing the servo set point from its initial constant value.

The early growth of the crystal, immediately after the nucleation pulse, is strongly affected by the Gibbs-Thomson effect, as the crystal size is exceedingly small. The size measurements are uncertain in this regime as well, owing to the finite resolution of the optical system. For these reasons, the model ignores the first two data points shown in Figure 6.

As described above, the zero-point voltage $V_{mon,0}$ was treated as an adjustable parameter in the model. The point at which $R(t)$ began to diminish was used to choose $V_{mon,0}$, but it was also quite useful to vary $V_{mon,0}$ to see how this affected the fit to the data, as a way to better understand the model uncertainties.

As seen in Figure 6, the model was abruptly terminated once $dR/dt$ became negative. This happened at a slightly positive supersaturation, owing to the Gibbs-Thomson effect. In principle, the model could be extended to describe the subsequent sublimation, but I have not gotten satisfactory results in this regime. During sublimation, the morphology changes from that of a faceted prismatic crystal to a rounded figure, and I have not yet developed the model to the point that it adequately reproduces the sublimation behavior.

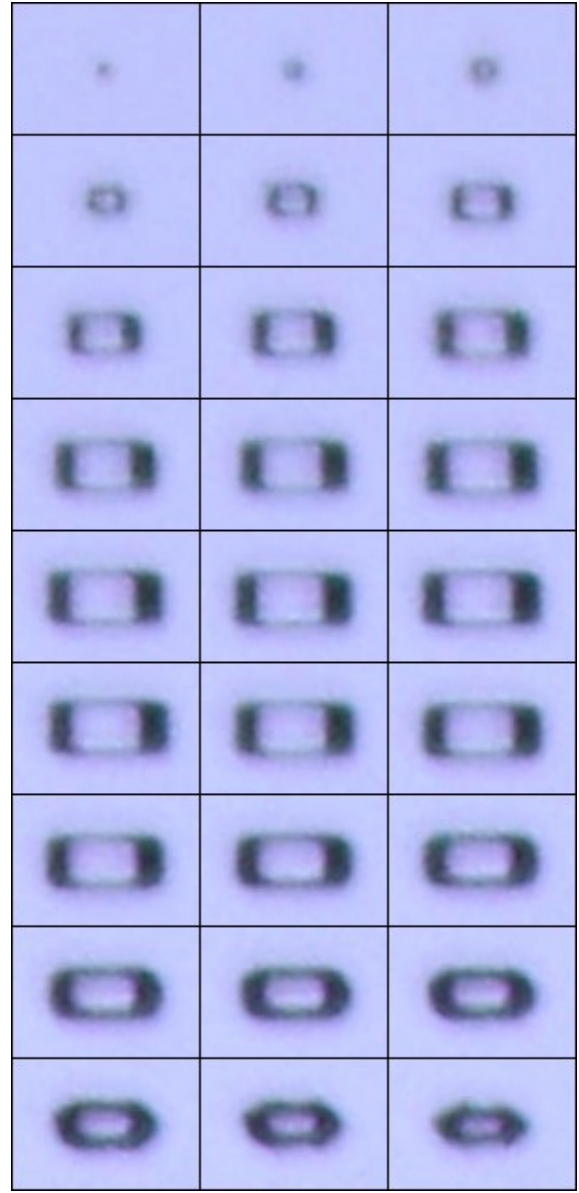

**Figure 5: Sub-images showing the growth and sublimation of the test crystal circled in Figure 4. The time series runs from the upper left to the lower right. The distance between the basal surfaces is 2H, while R denotes the effective radius of the hexagonal prism. Measurements from these images yielded the data points shown in Figure 6.**

The model shown in Figure 6 used $\alpha_{basal}(\sigma_{surf}) = \exp(\sigma_{0,basal}/\sigma_{surf})$ and $\alpha_{prism}(\sigma_{surf}) = A_{prism}\exp(\sigma_{0,prism}/\sigma_{surf})$ with $\sigma_{0,basal} = 0.73\%$, $\sigma_{0,prism} = 0.20\%$, and $A_{prism} = 0.25$. Additional results of this nature can be found in [2019Lib2]. Note that the basal growth slowed and then halted substantially sooner than the prism growth, reflecting the higher nucleation barrier on the basal surfaces.

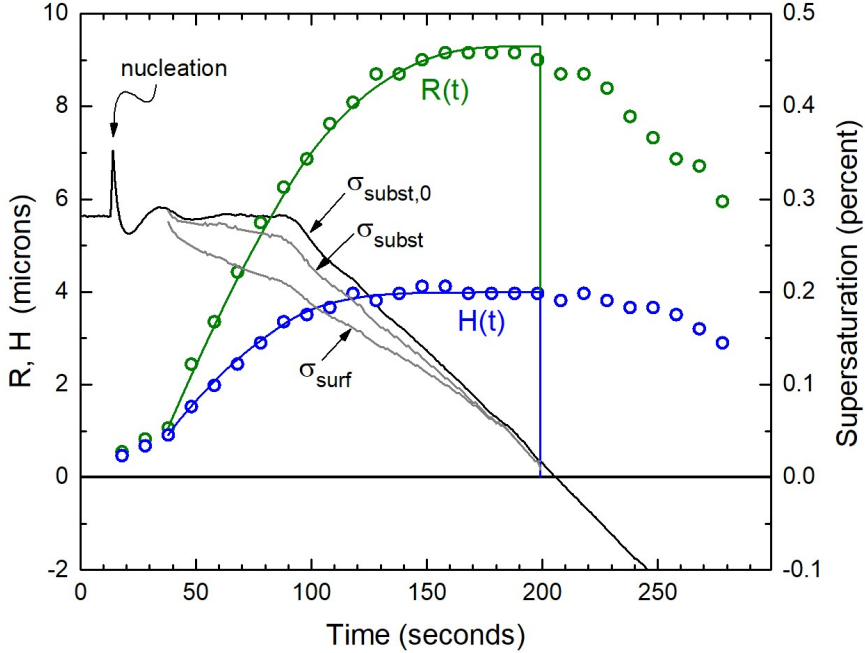

Figure 6: The green and blue data points show measurements obtained from the set of images shown in Figure 5. The green and blue lines are from the growth model described in the text. The heavy black line shows $\sigma_{subst,0}(t)$ determined from the measured $V_{mon}(t)$ together with the model parameter $V_{mon,0}$, as given in Equation 4. The grey line below $\sigma_{subst,0}(t)$ shows $\sigma_{subst}(t)$, while the line below that gives $\sigma_{surf}(t)$, both derived from the model.

Note also that the supersaturation corrections increased rapidly during the early growth, reflecting the rapidly increasing crystal size at this time. As the supersaturation decreased, the growth velocity decreased as well, yielding ever smaller supersaturation corrections during this time.

A particularly useful feature of this analysis model is its ability to examine systematic errors and uncertainties in both the measurements and analysis. For example, refitting the data (using different values of $\sigma_{0,basal}$, $\sigma_{0,prism}$, and other parameters) with different assumptions of $V_{mon,0}$ provides a good indication of how much the kinetic parameters change with the uncertainty in determining $V_{mon,0}$. By running many models using a variety of different input assumptions, it soon becomes fairly straightforward to extract a reasonable best-fit $\sigma_{0,basal} \pm \delta\sigma_{0,basal}$ (for example) that includes a realistic uncertainty in the measured value. Alternatively, as was done in [2019Lib2], the model runs can also be used to extract measurements of $\alpha_{basal}$ and $\alpha_{prism}$ as a function of $\sigma_{surf}$, along with a reasonable understanding of the measurement and modeling uncertainties. While this overall analysis process is necessarily somewhat subjective, in practice it works quite well, and I have found that this strategy of examining many forward-modeling runs using different parameters is substantially superior to other analysis options.

In summary, the apparatus described above allows for precise measurements of the growth of simple ice prisms over a broad range of environmental conditions, specifically as a function of temperature, supersaturation, and background gas pressure. The experimental methodology is especially well suited for observing small ice crystals, just a few microns in size, in order to reduce unwanted particle-diffusion effects. A versatile modeling strategy was also developed to disentangle the various physical processes that contribute to the observed crystal growth rates. This apparatus makes it possible to gain new insights into the ice/vapor attachment kinetics and, more generally, to better understand the fundamental physical processes that govern snow crystal growth.

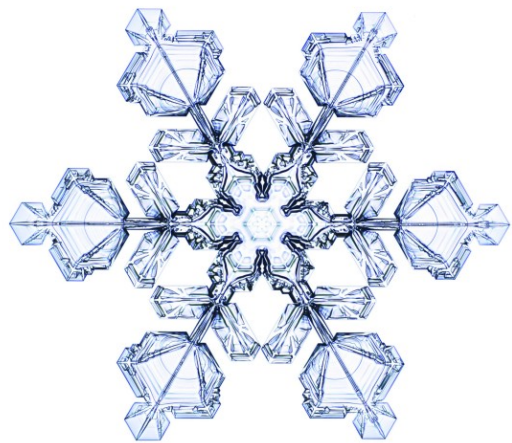

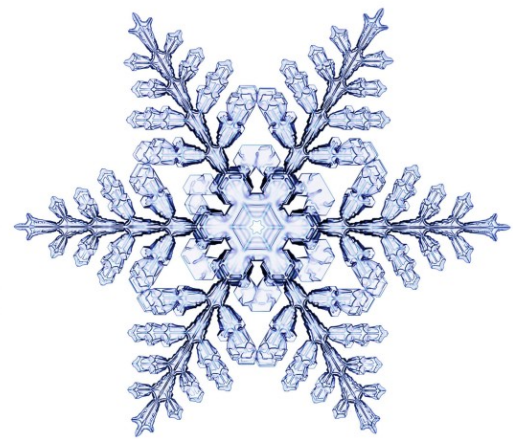